\documentclass[11pt]{article}
\usepackage[utf8]{inputenc}
\usepackage{booktabs}
\usepackage{hyperref}
\usepackage{epsfig,psfig}

\def\simgt{\lower 2pt \hbox{$\, \buildrel {\scriptstyle >}\over {\scriptstyle \sim}\,$}}
\def\simlt{\lower 2pt \hbox{$\, \buildrel {\scriptstyle <}\over {\scriptstyle \sim}\,$}}

\def\euclid{{\it Euclid\/}}
\def\herschel{{\it Herschel\/}}

\def\spitzer{{\it Spitzer\/}}
\def\wfirst{{\it WFIRST\/}}

\def\xray{{\hbox{X-ray}}}

\title{Active Galaxy Science in the LSST Deep-Drilling Fields: Footprints, Cadence Requirements, and Total-Depth Requirements}
\author{\small
W.N. Brandt (Penn State), 
Q. Ni (Penn State), 
G. Yang (Penn State),
\\ 
\small
S.F. Anderson (Univ Washington), 
R.J. Assef (Univ Diego Portales), 
A.J. Barth (UC Irvine),
\\
\small
F.E. Bauer (Cat\'olica),
A. Bongiorno (Oss Ast Roma),
C.-T. Chen (MSFC),
\\
\small
D. De~Cicco (Cat\'olica),
S. Gezari (Univ Maryland),
C.J. Grier (Penn State),
\\
\small
P.B. Hall (York Univ),
S.F. Hoenig (Univ Southampton),
M. Lacy (NRAO),
\\
\small
J. Li (Univ Illinois),
B. Luo (Nanjing Univ), 
M. Paolillo (Univ Naples Fed II), 
\\
\small
B.M. Peterson (Ohio State),
L.\v C. Popovi\'c (Ast Obs Belgrade),
G.T. Richards (Drexel Univ), 
\\
\small
O. Shemmer (Univ N Texas),
Y. Shen (Univ Illinois),
M. Sun (USTC), 
\\
\small
J.D. Timlin (Penn State),
J.R. Trump (Univ Connecticut),
F. Vito (Cat\'olica),
\\
\small
Z. Yu (Ohio State)}
\date{November 2018}

\begin{document}

\maketitle

\begin{abstract}
\noindent
This white paper specifies the footprints, cadence requirements, and
total-depth requirements needed to allow the most-successful AGN studies in 
the four currently selected LSST Deep-Drilling Fields 
(DDFs): \hbox{ELAIS-S1}, XMM-LSS, CDF-S, and COSMOS. 
The information provided on cadence and total-depth requirements will also likely 
be applicable to enabling effective AGN science in any additional DDFs 
that are chosen. 
\end{abstract}

\section{White Paper Information}

The contact author for this white paper is W.N. Brandt (wnbrandt@gmail.com). 

\bigskip

\noindent
This white paper addresses active galactic nucleus (AGN) science in the LSST Deep-Drilling Fields (DDFs),
including transient supermassive black hole (SMBH) activity. It is thus relevant to the ``Exploring the 
Changing Sky" main LSST science theme. 

\bigskip

\noindent
The survey type relevant to this white paper is the LSST DDFs. 

\bigskip

\noindent
This is an ``integrated program'' with science that hinges on the combination of pointing
and detailed observing strategy. 

\clearpage

\section{Scientific Motivation}

The aim of this white paper is to specify the footprints, cadence requirements, 
and total-depth requirements needed to allow successful AGN studies in 
the four currently selected LSST DDFs. 

\textit{\textbf{Footprints:}}
AGNs are fundamentally multiwavelength sources, and thus their effective 
selection and study are largely enabled by available high-quality multiwavelength 
data. The LSST DDFs were chosen to have among the best multiwavelength coverage 
in the sky, and substantial additional multiwavelength data were 
gathered in the DDFs after they were announced (see Figure~1). These DDF 
multiwavelength data can be used to assemble a ``ground-truth'' complete census of 
reliable AGNs (with sky density \hbox{1000--2000~deg$^{-2}$}), both obscured and 
unobscured, that is then used to train AGN selection and photometric-redshift
derivation in the main LSST survey. Below we propose precise footprints for 
the four currently selected LSST DDFs that optimize overlap with the available 
multiwavelength data. 

\textit{\textbf{Cadence Requirements:}}
We desire the DDFs to be observed at least every two nights in $grizy$ over the 
longest possible observing seasons during the full 10~yr LSST survey (additional 
key cadence details are given below). This cadence is required to provide the following: 

\begin{enumerate}

\item 
Photometric support for multi-object spectroscopic reverberation 
mapping (RM) campaigns, including the
SDSS-V Black Hole Mapper (BHM; Kollmeier et~al. 2018, arXiv:1711.03234) and 
the 4MOST TiDES-RM Project (S. Hoenig, private communication). These aim 
to derive broad-line region (BLR) properties and reliable SMBH masses for 
distant AGNs with expected observed-frame reverberation lags of \hbox{10--1000 days}.
{\it Dense, high-quality supporting photometry\/} from LSST will
substantially improve the accuracy/precision of derived SMBH masses
and also improve the number of reverberation-lag
detections, as demonstrated via both Monte Carlo simulations 
(e.g., Shen et al. 2015, ApJS, 216, 4; J. Li et~al., in preparation)
and analyses of present RM data (e.g., Grier et al. 2017, ApJ, 851, 21). 

\item
Dense, high-quality photometry {\it spanning multiple bands\/} that is powerful  
for studying accretion disks via continuum reverberation lags 
(e.g., Fausnaugh et~al. 2016, ApJ, 821, 56; Jiang et~al. 2017, ApJ, 836, 186; 
Mudd et~al. 2018, ApJ, 862, 123; Homayouni et~al., arXiv:1806.08360) 
for $\approx 3000$ AGNs spanning a wide range 
of luminosity and redshift. The proposed two-night sampling is critical to ensure 
a high lag recovery fraction for the relatively short accretion-disk lags expected
(Yu et~al., arXiv:1811.03638), and observations spanning 10~yr will allow 
investigations of secular evolution of these lags that test the underlying model. 
Basic RM for the BLR entirely based on photometric data may also be 
possible (e.g., Chelouche et~al. 2014, ApJ, 785, 140).

\item 
Data for calibrating long-term AGN variability selection using the
\hbox{40,000--80,000} AGNs in the ground-truth sample
(e.g., De~Cicco et~al. 2015, A\&A, 574, 112; 
Falocco et~al. 2015, A\&A, 579, 115). 
Here it is essential to reach the long timescales where the red-noise 
power spectral density function indicates that AGN variability will 
be strongest, and mild long-term changes in obscuration/scattering may 
allow even obscured AGNs to be identified with the high LSST signal-to-noise.  

\item 
Critical tests of general AGN continuum variability, as described by 
damped random walk and other models (e.g., MacLeod et~al. 2012, ApJ, 753, 106; 
Kasliwal et~al. 2015, MNRAS, 451, 4328). 
Well-calibrated continuum-variability models, including both flux
and color information, will be essential for making optimum use of 
AGN variability information in the main LSST survey. 
Exceptional modes of AGN variability will also be investigated, 
including color-dependent AGN periodicity or quasi-periodicity
on week-to-year timescales from, e.g., 
accretion-disk ``hot spots'', 
density waves, or 
close binary SMBHs
(e.g., Graham et~al. 2015, MNRAS, 453, 1562; 
Bhatta et~al. 2016, ApJ, 832, 47;
Charisi et~al. 2018, MNRAS, 476, 4617; 
Smith et al. 2018, ApJL, 860, L10). 

\item 
Exploration of transient phenomena such as 
stellar tidal disruptions (e.g., Stone et~al. 2018, arXiv:1801.10180), 
changing-look AGNs, 
jet-linked flares, and 
intermediate-mass black hole outbursts. 
Such transient behavior arises over 
timescales spanning days to years. 

\end{enumerate}

\textit{\textbf{Total-Depth Requirements:}} 
We desire total depths of $z\approx 27.9$ and $y\approx 26.5$ to detect 
statistically meaningful samples of $\approx 30$ $(\approx 100)$ AGNs at 
$z\simgt 7$ $(z\simgt 6)$; e.g., see Matsuoka et~al. (2018, arXiv:1811.01963). 
The total depths of $g$, $r$, and $i$ will 
be \hbox{28.3--28.7} after satisfying the 
AGN variability cadence requirements. We require the total 
depth in $u$ also reach $\approx 28.3$ to effectively improve 
photometric-redshift quality (especially in reducing 
outliers) for AGNs detected in $g$, $r$, and $i$ 
at $z\simlt 4$. These deep optical data will also help 
determine the nature of optically faint AGNs identified at 
radio, \xray, and other wavelengths 
(e.g., Richards et~al. 1999, ApJ, 526, 73; 
Mainieri et~al. 2005, A\&A, 437, 805; 
Luo et~al. 2010, ApJS, 187, 560).

\vspace{.6in}

\section{Technical Description}


\subsection{High-level description}

Below we specify the footprints, cadence requirements, and
total-depth requirements needed to allow the most-successful AGN studies in 
the four currently selected LSST DDFs: ELAIS-S1, XMM-LSS, CDF-S, and COSMOS. 

\vspace{.3in}

\subsection{Footprints -- pointings, regions, and/or constraints}

The LSST field of view is sufficient to cover the prime 
multiwavelength data available in each of the four DDFs, so
we request a single pointing position for each of the DDFs.
To optimize overlap with the available multiwavelength data, 
we propose that the four LSST DDFs be centered at the 
positions listed in the table in this section. 

\begin{table}[h!]
\begin{tabular}{|c|c|c|}
\hline
Field   & Central RA  & Central Dec   \\
Name    & (J2000)     & (J2000)       \\ 
\hline  
ELAIS-S1  & 00:37:48   & $-$44:01:30 \\
XMM-LSS   & 02:22:18   & $-$04:49:00 \\
CDF-S     & 03:31:55   & $-$28:07:00 \\
COSMOS    & 10:00:26   & $+$02:14:01 \\
\hline
\end{tabular}
%
%
\end{table}

\noindent
In \hbox{Figures~2--5}, we show how the resulting LSST fields align versus 
the available multiwavelength data. 
In choosing the field central positions, we have prioritized
ensuring overlap with the sky regions covered by the 
SERVS (Mauduit et~al. 2012, PASP, 124, 714), 
5~Ms XMM-SERVS (e.g., Chen et~al. 2018, MNRAS, 478, 2132), 
VIDEO (Jarvis et~al. 2013, MNRAS, 428, 1281), and 
\spitzer\ DEEPDRILL (M. Lacy et~al., in preparation) programs; 
the SERVS, XMM-SERVS, and VIDEO coverage is closely 
overlapping. Additional deep \spitzer\ coverage to be 
obtained via the ``Cosmic Dawn Survey'' 
(PI: P. Capak; Program 13058) should lie within
the LSST CDF-S field, and
additional \spitzer\ coverage to be 
obtained via the ``Missing Piece'' survey 
(PI: A. Sajina; Program 14081) should lie within
the LSST COSMOS field. 
Deep \herschel\ coverage is also deemed essential, since it 
is needed for measurements of star formation in AGN hosts and 
will be irreplaceable in the near future. 

Small ($\simlt 5^\prime$) observation-to-observation offsets are 
acceptable to accommodate dithering (which is desired). The
rotational dithering plan put forward by the Dark Energy 
Science Collaboration (DESC) will likely
be acceptable for our AGN studies. 

We note that the XMM-LSS field is located near the bright ($V=6.5$) 
star Mira (located at $\alpha_{2000}=02$:19:20.8 and 
$\delta_{2000}=-02$:58:39.5). Mira lies 1.98~deg from the 
XMM-LSS field center listed above, and thus it will lie
just outside the LSST field of view (see Figure~6). An 
analysis of possible optical artifacts due to Mira is 
needed by LSST imaging-system experts. If required, the 
central pointing position for the XMM-LSS field could be 
shifted southward by up to 0.2~deg without significant loss of 
overlap with extant multiwavelength data. Additional southward 
shifting may also be possible if necessary, likely combined 
with some shifting in right ascension as well. 

The multi-object RM projects described above will utilize wide-field
spectrographs with fields that fit within the LSST field size. 
For reference, the SDSS-V Apache Point Observatory (APO) and 
Las Campanas Observatory (LCO) spectrograph field sizes are 
$\approx 6$~deg$^2$ and $\approx 2.6$~deg$^2$, respectively. 
The SDSS-V RM team is considering performing multiple spectroscopic 
pointings within the DDFs observed at LCO in order to fill the
DDFs more completely. 
The 4MOST spectrograph field size is $\approx 3.5$~deg$^2$. 
Other spectrographs that will likely be used for 
intensive DDF follow-up studies (e.g., Subaru PFS and VLT MOONS)
fit well within the LSST field size. 

Our focus in this white paper is on the four currently selected
LSST DDFs, so we have not suggested footprints for new DDFs. 
This being said, we desire to perform AGN studies in 
any other suitable new DDFs that are selected, such as the
Akari Deep Field-South being proposed by the DESC; this field
will have excellent synergy with \euclid\ and \wfirst. 
We would like to have the opportunity to provide constructive 
input on new DDF footprints, cadences, and other properties from 
the perspective of AGN investigations.  

\subsection{Image quality}

We request that the delivered $r$-band image quality be better than 
$1.2^{\prime\prime}$, with other filters to be scaled accordingly. This
should be achieved by LSST on most observing nights 
(e.g., see Figure~1 of Ivezi\'c et~al. 2018, arXiv:0805.2366
and associated discussion). Indeed, the median $r$-band free-air seeing 
at Cerro Pach\'on is $0.65^{\prime\prime}$. We therefore expect to 
have many observations with superb seeing that can be co-added for 
optimal AGN host-galaxy studies. 

\subsection{Individual image depth and/or sky brightness}

To optimize photometric-redshift derivation and source characterization
for AGNs and galaxies in the DDFs, we would like to achieve a relatively 
uniform depth across the LSST filters (see Chapters~3 and 10 of the 
LSST Science Book). This is economically possible for 
$ugri$ but less so for $z$ and $y$. We thus
request, every two nights, 
1 visit in $g$, 
1 visit in $r$, 
3 visits in $i$, 
5 visits in $z$, and  
4 visits in $y$. 
Here each visit is the standard 30~s. The 
3 visits in $i$ can be back-to-back
for sake of efficiency, as can the 
5 visits in $z$ and  
4 visits in $y$.
The $u$-band is addressed below. The following table summarizes 
the desired visits (depths quoted are $5\sigma$ design-specification 
depths): 

\begin{table}[h!]
\begin{tabular}{|l|c|c|c|c|c|c|c|}
\hline
Quantity of Interest   & $u$    & $g$    & $r$    & $i$    & $z$  & $y$   \\ 
\hline  
Visits Every 2 Nights  & 4      & 1      & 1      & 3      & 5    & 4     \\
Depth Every 2 Nights   & 24.6   & 25.0   & 24.7   & 24.6   & 24.2 & 22.9  \\
\hline
Total Visits in 10~yr  & 3600   & 900    & 900    & 2700   & 4500 & 3600  \\
Total Depth in 10~yr   & 28.3   & 28.7   & 28.4   & 28.3   & 27.9 & 26.5  \\
\hline
\end{tabular}
%
%
\end{table}

The $grizy$ visits in a given night should be obtained as close in 
time as reasonably possible to minimize any intranight variability effects. 
If it can be accommodated, the order of priority for observations 
should be $g$, then $i$, then $r$, $z$, and $y$. The bluest filter 
$(g)$ is the most important for monitoring continuum variations; 
$g-i$ is the most widely useful color for AGN studies; and the 
remaining filters are ordered by their expected total depth per 
night. The priority with which bands are to be observed is relevant 
because observations potentially can be interrupted at any time due 
to, e.g., environmental conditions crossing a threshold that halts 
observing in the current field, or unexpected software or 
hardware faults. 

The absolute number of visits every two nights 
is set by our requirements for total survey depth as well
as our coordination with the DESC (see \S3.12 for more details). 
The DESC wishes to obtain deep $grizy$ imaging 
with a three-night cadence, primarily to allow study of 
faint, distant supernovae. Indeed, they desire 
even deeper $griz$ imaging 
than we request. We can accommodate 
such deep imaging provided it does not lead 
to a reduction in the observing frequency from our desired
two-night cadence, and it will provide outstanding 
photometric signal-to-noise for AGN variability studies. 

Observations in the $u$-band are critical for AGN photometric-redshift
derivation and AGN physics studies, and thus we desire to obtain 
deep $u$ coverage in the DDFs as well. We recognize
that this coverage likely cannot be obtained in the same 
synoptic manner requested for the other bands 
since the $u$ filter will often not be in 
the filter wheel and $u$ observations will be concentrated 
during dark time to improve efficiency. We request 60 visits per
month in $u$, and we would like these to have the best
synoptic coverage that general LSST operations can
reasonably allow (each $u$ synoptic observation would ideally 
have 4 consecutive visits in $u$ to match the depths of the 
other bands). When the $u$ filter is available, it should have 
the highest priority for observation to ensure that an 
adequate number of visits are obtained with that filter.

The $u$-band is also critical for distinguishing the tidal 
disruption and accretion of a star by a SMBH (a tidal disruption 
event) from more common extragalactic transients such as supernovae.
In particular, the $u-r$ color and color evolution with time can 
effectively remove most supernovae that happen to be coincident 
with the nuclei of galaxies (e.g., van~Velzen et~al. 2011, ApJ, 741, 73).
Given the slow rise times of tidal disruption events of \hbox{1--2} 
months (e.g., van~Velzen et~al. 2018, arXiv:1809.02608), $u$-band coverage with a 
cadence of \hbox{1--2} weeks would be sufficient to confirm the 
persistent blue nature of a transient on its rise to peak and beyond.
Similarly, $u$-band coverage is valuable in searching for the 
expected relativistic Doppler boosting associated with gas 
bound to the individual SMBHs in close-separation SMBH binaries 
(e.g., Charisi et~al. 2018). 

\subsection{Co-added image depth and/or total number of visits}

With observing-season lengths of \hbox{7--8.5} months annually, and 
weather losses of 25\%, we expect to obtain the following 
approximate total numbers of visits over the 10~yr LSST survey:
3600 visits in $u$, 
900 visits in $g$, 
900 visits in $r$, 
2700 visits in $i$, 
4500 visits in $z$, and  
3600 visits in $y$. 
These are expected to reach co-added depths of 
28.3 mag in $u$, 
28.7 mag in $g$, 
28.4 mag in $r$, 
28.3 mag in $i$, 
27.9 mag in $z$, and  
26.5 mag in $y$. 
As desired, the depths are relatively uniform in $ugri$. 
In $z$ and $y$ the depths are unavoidably somewhat shallower, 
but they have been made as deep as economically possible. 
Achieving high sensitivity in $z$ and $y$ is critically
important, despite the relatively high expense; e.g., 
for selecting and studying high-redshift AGNs. 


\subsection{Number of visits within a night}

This topic is addressed in \S3.4.

\subsection{Distribution of visits over time}

Our time-domain science described in \S2
requires observations at least every two nights in $grizy$. 
We desire observations of each field spaced every other night throughout 
their observing seasons. To mitigate the effects of weather or other 
losses, if no observations (or incomplete observations) are made in a 
given night then we would request elevated priority for the next 
available night in order to make up the loss if possible, after which 
we would resume observations on the originally planned nights. This 
approach would yield some observations separated by one night instead 
of two; such observations will be useful for AGN variability studies 
on shorter timescales. 

RM studies benefit from the {\it longest observing seasons possible\/}; this 
increases the chances of robust lag detection, allows access to longer lags 
corresponding to higher SMBH masses, and minimizes missed lags due to 
non-overlapping data in cross-correlation analyses. 
We therefore do not want the DDF observations
by LSST to be tightly clustered in a narrow time window, as has been done
in some recent LSST operations simulations (to minimize typical
airmass). The current planning for SDSS-V calls for its RM fields to 
be observed spectroscopically for \hbox{6--7} months each year, and 
we expect similar windows for \hbox{4MOST} RM. We request that the 
LSST observations span, at least, this same time window 
plus \hbox{1--1.5} month precursor observations every season before the 
spectroscopic observations begin. The precursor observations will allow the 
``driving'' AGN continuum to be sampled earlier than 
the ``responding'' BLR emission. We can tolerate airmass values 
up to $\approx 2.0$ in order to achieve these long \hbox{(7--8.5 month)}
LSST observing seasons. We are aware that the DESC also desires
the longest LSST observing seasons possible in order to optimize their 
studies of distant supernovae, and we expect that a mutually agreeable
solution can be found.  

The search for SMBH binaries will also benefit from the longest 
observing seasons possible. Discriminating true periodicity from 
an accreting SMBH binary vs.\ red noise associated with a normal 
AGN (powered by a single accreting SMBH) requires observing multiple 
cycles of variation. Specifically, three or more well-sampled cycles are
required to reduce the false-alarm probability, as shown from 
simulations (Vaughan et~al. 2016, MNRAS, 461, 3145) and extended 
baseline monitoring of reported SMBH binary 
candidates (Liu et~al. 2018, ApJ, 859, 12). Longer observing
windows will allow LSST to detect larger ranges of periods robustly
within its 10~yr baseline. The periods of interest for which a 
SMBH binary ($10^7$--$10^9$~M$_\odot$) is in the gravitational-wave 
driven regime of orbital decay span a week to a few years, easily 
probed by the DDFs if observed over the widest observing seasons 
possible. Also, given that the residence time for SMBH binaries increases 
for longer orbital periods (e.g., Haiman et~al. 2009, ApJ, 700, 1952), 
the probability of detecting a SMBH binary increases as longer
periods are probed. 

The SDSS-V BHM will operate from \hbox{2020--2025} (Kollmeier et~al. 2018) 
and the 4MOST TiDES-RM Project will operate from \hbox{$\approx 2021$--2026} 
with a possible additional five-year 
extension (S. Hoenig, private communication).\footnote{We also 
note that the Maunakea Spectroscopic Explorer (MSE) aims to perform RM in 
the near-equatorial DDFs with observations perhaps beginning in $\approx 2026$
(e.g., Flagey et al. 2018, arXiv:1807.08019).} 
SDSS-V BHM will target the XMM-LSS, CDF-S, and COSMOS DDFs, and
it aims to deliver $\approx 300$ SMBH masses in these fields.  
4MOST TiDES-RM will target all four DDFs, and it aims to deliver
$\approx 700$ SMBH masses in these fields. 
Both of these campaigns are expected to be ongoing at the start
of LSST full science operations in \hbox{2022--2023}, so LSST synoptic 
observations of the DDFs should begin promptly. If possible, we also strongly
desire LSST synoptic observations of the DDFs during the LSST Science
Verification period in \hbox{2021--2022}. The DDFs are among the best 
testing fields during the Science Verification period, which can be used 
to test the stability and performance of LSST as functions of time and 
observing conditions (e.g., seeing and airmass) for the same set of 
well-characterized sources. Doing so will also benefit time-domain 
science greatly by extending the effective total time baseline 
and delivering early time-domain data.


\subsection{Filter choice}

Our filter constraints are described above. 

\subsection{Exposure constraints}

We request standard 30~s exposures per visit. 
We desire to avoid saturation for AGNs as bright as $i=16$, and
this should be possible according to Section~3.2 of the LSST 
Science Book. 

\subsection{Other constraints}

Our constraints are suitably described above. 

\subsection{Estimated time requirement}


We can estimate the fraction of the total LSST time that we request be
dedicated to this program. With $\approx 7.5$ month observing seasons, 
we will have $\approx 112$ observing nights each season in the absence
of weather losses. Including 25\% weather losses, this number will be 
reduced to $\approx 84$. Each of these
will utilize 18 visits, following \S3.4 (the $u$-band observations will
be parceled out differently, but we have used the appropriate average
value here). Thus, annually each DDF will
require 1512 visits. For 10~yr and four DDFs, the total number of
DDF visits will be 60480. The total number of visits of LSST is $\approx 2.8$
million, and thus we request $\approx 2.2$\% of the total LSST time. 

A more accurate estimate of the fraction of total time required will come from the 
full requested simulation of this program. For example, this will allow
a more reliable assessment of weather losses.

\vspace{.3in}

\begin{table}[ht]
    \centering
    \begin{tabular}{l|l|l|l}
        \toprule
        Properties & Importance \hspace{.3in} \\
        \midrule
        Image quality &    2 \\
        Sky brightness &   2  \\
        Individual image depth & 1  \\
        Co-added image depth &  1  \\
        Number of exposures in a visit   &  2  \\
        Number of visits (in a night)  &  1 \\ 
        Total number of visits &    1 \\
        Time between visits (in a night) & 2  \\
        Time between visits (between nights)  & 1   \\
        Long-term gaps between visits &  1 \\
        Other (please add other constraints as needed) & \\
        \bottomrule
    \end{tabular}
    \caption{{\bf Constraint Rankings:} Summary of the relative importance of various survey strategy constraints. Please rank the importance of each of these considerations, from 1=very important, 2=somewhat important, 3=not important. If a given constraint depends on other parameters in the table, but these other parameters are not important in themselves, please only mark the final constraint as important. For example, individual image depth depends on image quality, sky brightness, and number of exposures in a visit; if your science depends on the individual image depth but not directly on the other parameters, individual image depth would be `1' and the other parameters could be marked as `3', giving us the most flexibility when determining the composition of a visit, for example.}
        \label{tab:obs_constraints}
\end{table}

\subsection{Technical trades}

\noindent
{\it What is the effect of a trade-off between your requested survey 
footprint (area) and requested co-added depth or number of visits?\/}

\vspace{0.05 in}

\noindent
Not applicable. This white paper is focused upon the four 
LSST Deep-Drilling Fields that have already been selected: 
\hbox{ELAIS-S1}, XMM-LSS, \hbox{CDF-S}, and COSMOS. Thus, the total 
solid-angle coverage on the sky is already largely fixed. 

\vspace{0.2 in}


\noindent
{\it If not requesting a specific timing of visits, what is the effect 
of a trade-off between the uniformity of observations and the 
frequency of observations in time? e.g. a `rolling cadence' 
increases the frequency of visits during a short time period 
at the cost of fewer visits the rest of the time, making 
the overall sampling less uniform.\/}

\vspace{0.05 in}

\noindent
We {\it are\/} requesting a specific timing of visits. We do not
want a ``rolling cadence'' in the DDFs. 

\vspace{0.2 in}


\noindent
{\it What is the effect of a trade-off on the exposure time and 
number of visits (e.g. increasing the individual image 
depth but decreasing the overall number of visits)?\/}

\vspace{0.05 in}

\noindent
Decreasing the overall number of visits would have a 
substantial negative effect upon this program. Even a
minor reduction in the overall number of visits 
(e.g., from a two-night cadence to a three-or-four-night cadence)
would damage the relatively rapid time-domain science 
including 
continuum RM studies of accretion disks (point \#2 in \S2),  
general/exceptional AGN variability studies (point \#4 in \S2), and 
exploration of transient SMBH phenomena (point \#5 in \S2). 
A larger reduction in the overall number of visits would
damage the program even more broadly; e.g., harming  
photometric support of multi-object spectroscopic RM campaigns (point \#1 in \S2). 

\vspace{0.2 in}


\noindent
{\it What is the effect of a trade-off between uniformity 
in number of visits and co-added depth? Is there any 
benefit to real-time exposure time optimization to 
obtain nearly constant single-visit limiting depth?\/}

\vspace{0.05 in}

\noindent
We do not want a reduction in the number of visits or a 
reduction in the uniform two-night cadence. We do desire
a relatively constant single-visit depth, but our 
constraints here are not highly demanding. For example, 
it would be helpful to implement a first-order correction
for expected light loss when the airmass is high. Single-visit
depths that are constant to within 0.2 mag should be acceptable. 

\vspace{0.2 in}


\noindent
{\it Are there any other potential trade-offs to consider 
when attempting to balance this proposal with others 
which may have similar but slightly different requests?\/}

\vspace{0.05 in}

\noindent
We have coordinated significantly with the DESC in designing
the cadence proposed here; this productive coordination started 
at the LSST Cadence Hackathon at the Flatiron Institute in 
2018 September. While we have made much progress in 
converging upon a cadence solution that is agreeable 
to both Science Collaborations, there are still some
remaining tensions (particularly regarding a 
two-night vs.\ three-night cadence). 
The DESC wishes to obtain deep $grizy$ imaging every three nights, 
primarily to allow study of faint, distant supernovae; indeed, 
they desire even deeper $griz$ imaging 
than we request. We can accommodate 
such deep imaging {\it provided it does not lead 
to a reduction in the observing frequency from our desired
two-night cadence\/}, and it will provide outstanding 
photometric signal-to-noise for AGN variability studies. 
Both Science Collaborations share a strong preference for
the longest LSST observing seasons practically possible. 

\vspace{0.2 in}


\clearpage

\section{Performance Evaluation}

The following heuristics tied to observing strategy can be used
for evaluation of performance: 

\begin{enumerate}

\item 
We require that the LSST DDFs {\it cover the prime multiwavelength data 
available in each of the four fields\/}, as illustrated in 
\hbox{Figures~1--5}. In choosing the field central positions
provided in \S3.2, we have prioritized
ensuring overlap with the sky regions covered by the 
SERVS, XMM-SERVS, VIDEO, and 
\spitzer\ DEEPDRILL programs; 
the SERVS, XMM-SERVS, and VIDEO coverage is closely 
overlapping. Deep \herschel\ coverage is also deemed essential, 
since it is needed for measurements of star formation in AGN 
hosts and will be irreplaceable in the near future. 
A relevant existing metric is 
``NightPointingMetric''.

\item
We require the DDFs to be observed {\it at least every two nights in 
$grizy$ in as regular a manner as possible\/}. 
Many additional relevant details, including $u$-band 
requirements, are provided in \S3.4. {\it Multiple-band coverage\/} is
especially required for photometric RM of AGN accretion disks, 
general AGN variability characterization, and SMBH transient 
characterization (see \S2). Longer sampling timescales 
would have a substantial negative effect upon this program. Even a
minor increase in the sampling timescale 
(e.g., from two nights to three-or-four nights)
would damage the rapid time-domain science 
including 
continuum RM studies of accretion disks (point \#2 in \S2),  
general/exceptional AGN variability studies (point \#4 in \S2), and 
exploration of transient SMBH phenomena (point \#5 in \S2). 
A larger increase in the sampling timescale would
damage the program even more broadly; e.g., harming  
photometric support of multi-object spectroscopic RM campaigns (point \#1 in \S2). 
Relevant existing metrics include 
``InterNightGapsMetric'', 
``LongGapAGNMetric'', and 
``MeanNightSeparationMetric''.

\item
To optimize photometric-redshift derivation and source characterization
for AGNs and galaxies in the DDFs, we would like to achieve a {\it relatively 
uniform depth across the LSST filters.\/} This is economically possible for 
$ugri$ but less so for $z$ and $y$. We thus
request, every two nights, 
1 visit in $g$, 
1 visit in $r$, 
3 visits in $i$, 
5 visits in $z$, and  
4 visits in $y$. 
Here each visit is the standard 30~s. The 
3 visits in $i$ can be back-to-back
for sake of efficiency, as can the 
5 visits in $z$ and  
4 visits in $y$.
The $u$-band is addressed in \S3.4. 
This pattern of visits across the LSST bands has been developed
with some coordination with the requirements of the DESC, and it will
satisfy our desires for total $ugrizy$ depths described in \S1. 
Relevant existing metrics include 
``AccumulateCountMetric'', 
``AccumulateM5Metric'', 
``Coaddm5Metric'', 
``CrowdingMagUncertMetric'',
``HistogramM5Metric'', and 
``NVisitsPerNightMetric''.

\item
We require the {\it longest observing seasons possible\/}. For example, 
this increases the chances of robust RM lag detection, allows access 
to longer RM lags corresponding to higher SMBH masses, minimizes missed 
RM lags due to non-overlapping data in cross-correlation analyses, and 
aids searches for SMBH binaries. The current planning for SDSS-V calls 
for its RM fields to be observed spectroscopically for \hbox{6--7} months 
each year, and we expect similar windows for \hbox{4MOST} RM. We request 
that the LSST observations span, at least, this same time window 
plus \hbox{1--1.5} month precursor observations every season before the 
spectroscopic observations begin. The precursor observations will allow the 
``driving'' AGN continuum to be sampled earlier than 
the ``responding'' BLR emission. We can tolerate airmass values 
up to $\approx 2.0$ in order to achieve these long \hbox{(7--8.5 month)}
LSST observing seasons.
A relevant existing metric is 
``CampaignLengthMetric''.

\item
We request that the delivered $r$-band image quality be better than 
$1.2^{\prime\prime}$, with other filters to be scaled accordingly.

\end{enumerate}


\vspace{.6in}

\clearpage

\section{Special Data Processing}


Below we provide a list of our main special data processing
requests. We share many of the same desires as the DESC. 

\begin{enumerate}

\item
A pipeline that creates single-night DDF co-adds in each filter is 
recommended for AGN variability and SMBH transient studies. This
pipeline should also perform difference-imaging analysis (DIA), and 
create DIASource and DIAObject catalogs for the DDFs. 
These images and catalogs should be updated and made available 
via the Science Platform as promptly as possible, before the start
of the next observing night. 

\item 
Alert packets should be created from the DDF DIASources and provided
to the community efficiently. 

\item
As described in \S3.7, observations of the DDFs are desired during
the LSST Science Verification period. These should be co-added to
create deep template images for the DDFs, in order to allow 
effective DIA for the DDFs during the first year of LSST
operations.  

\item
Annually, we request that weekly, monthly, and yearly 
co-adds in each filter for each DDF be created. These will 
allow, e.g., variability studies at fainter levels than are
possible in the nightly co-adds. These should be used to 
create DIASource and DIAObject catalogs for the DDFs reaching
the faintest flux levels possible.

\item 
Annually, we request that co-adds be created in each filter from 
the 10\%, 20\%, 30\%, 40\%, 50\%, 60\%, 70\%, 80\%, and 90\% 
of observations with the best imaging quality. These can be
used for optimal studies of AGN host galaxies. 

\item 
As described in \S3.2, there is a bright star, Mira, lying close 
to the XMM-LSS field. An analysis of possible optical artifacts
and data-processing issues due to Mira is needed by experts in
the LSST Project. Data processing should be performed in a manner 
that minimizes problems in science analysis due to Mira. 

\end{enumerate}



\clearpage

\begin{figure}[t!]
\vskip -0.10truein
\hbox{
\hskip +0.0in 
\psfig{figure=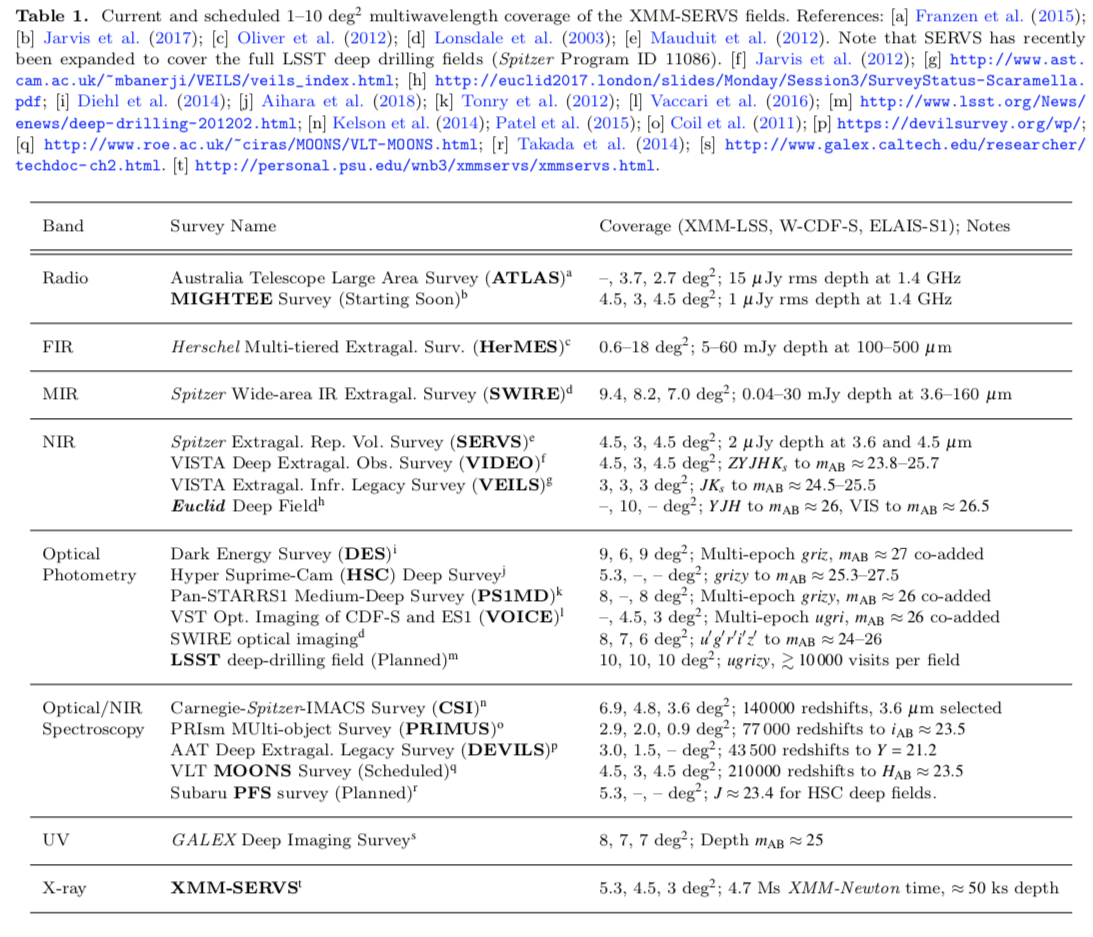,height=5.0truein,width=5.0truein,angle=0}
}
\vspace*{0.0in}
\caption[]{\protect\small 
Current and scheduled \hbox{1--10~deg$^2$} multiwavelength coverage for 
three of the LSST DDFs. From Chen et~al. (2018).}
\vspace*{-0.2in}
\end{figure}


\clearpage

\begin{figure}[t!]
\vskip -0.10truein 
\psfig{figure=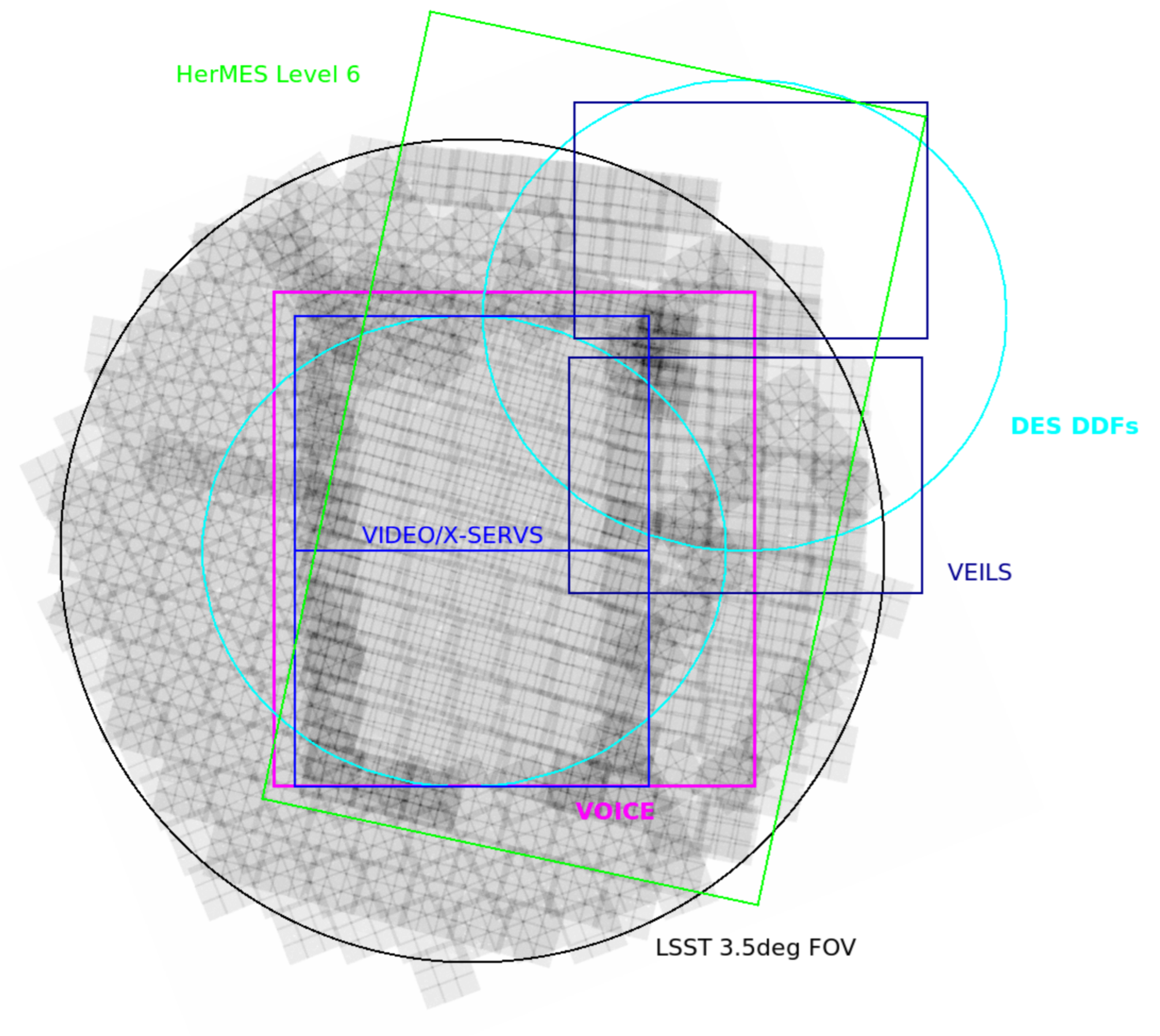,height=4.0truein,width=5.0truein,angle=0}
\vspace*{0.0in}
\caption[]{\protect\small 
Comparison of the planned LSST field vs.\ some of the key available 
multiwavelength data (as labeled) for ELAIS-S1. The background grayscale
image shows the available \spitzer\ 3.6~$\mu$m data.}
\vspace*{-0.2in}
\end{figure}


\clearpage

\begin{figure}[t!]
\vskip -0.10truein 
\psfig{figure=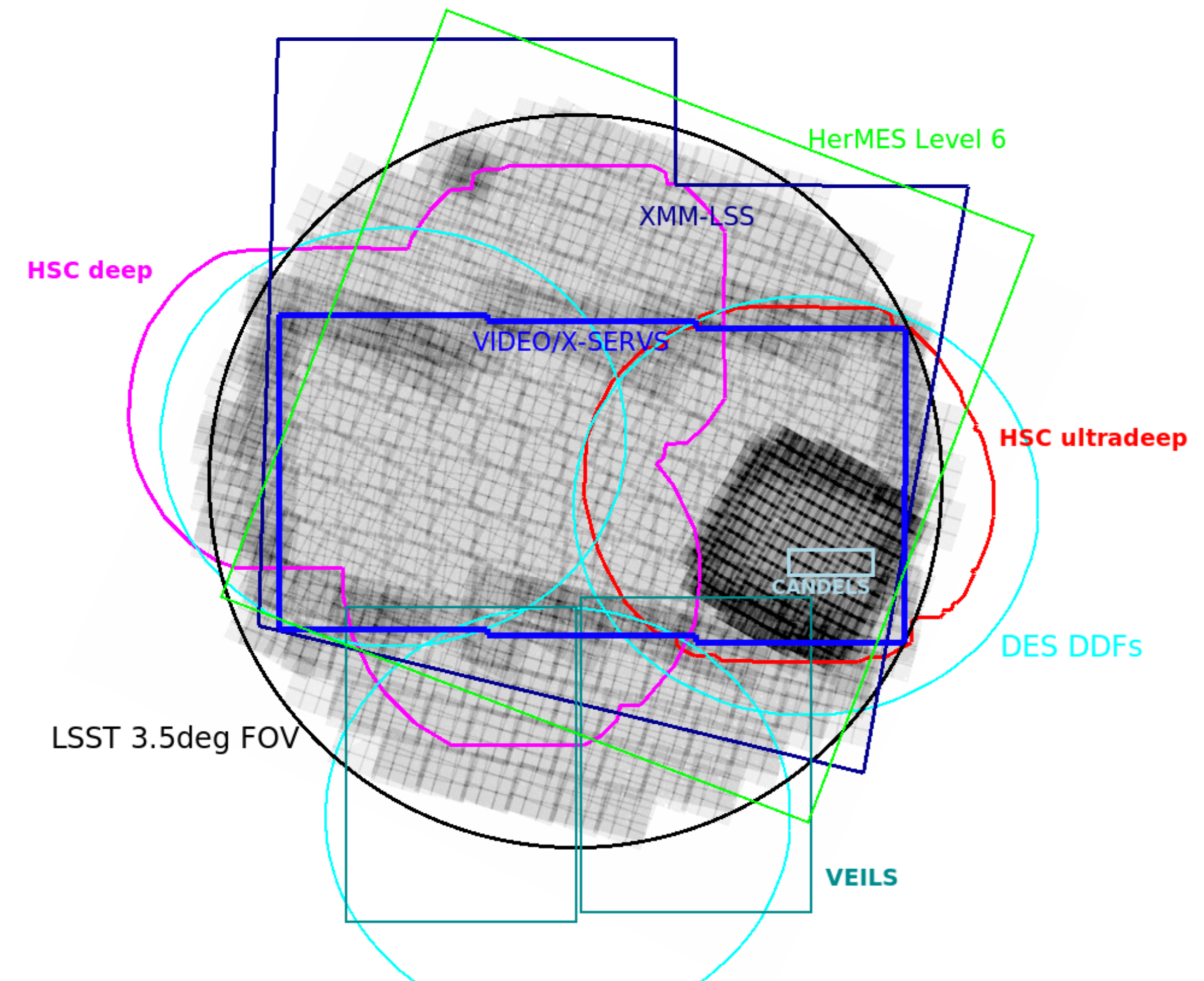,height=4.0truein,width=5.0truein,angle=0}
\vspace*{0.0in}
\caption[]{\protect\small 
Comparison of the planned LSST field vs.\ some of the key available 
multiwavelength data (as labeled) for XMM-LSS. The background grayscale
image shows the available \spitzer\ 3.6~$\mu$m data.}
\vspace*{-0.2in}
\end{figure}


\clearpage

\begin{figure}[t!]
\vskip -0.10truein 
\psfig{figure=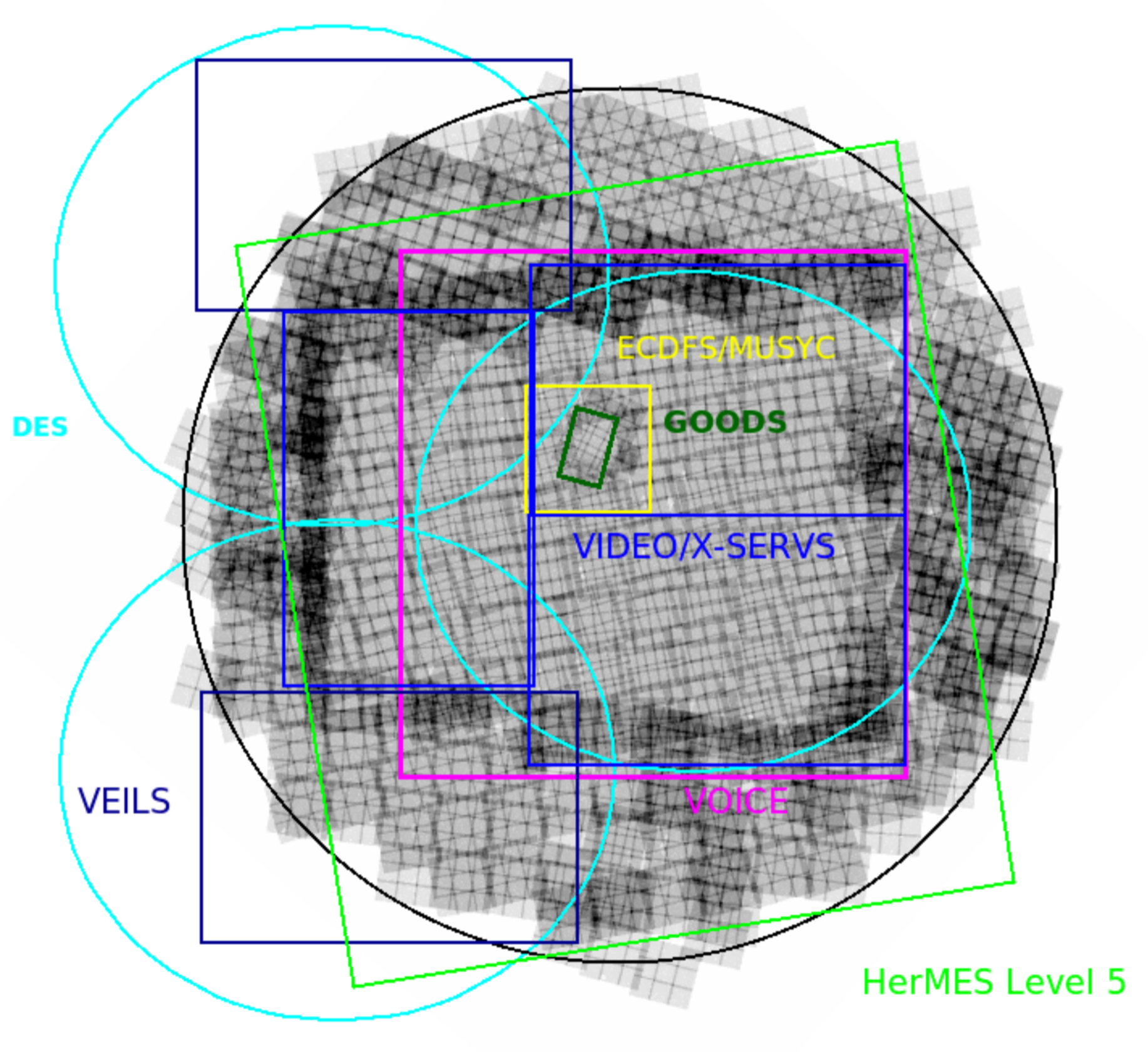,height=4.0truein,width=5.0truein,angle=0}
\vspace*{0.0in}
\caption[]{\protect\small 
Comparison of the planned LSST field vs.\ some of the key available 
multiwavelength data (as labeled) for CDF-S. The background grayscale
image shows the available \spitzer\ 3.6~$\mu$m data.}
\vspace*{-0.2in}
\end{figure}


\clearpage

\begin{figure}[t!]
\vskip -0.10truein 
\psfig{figure=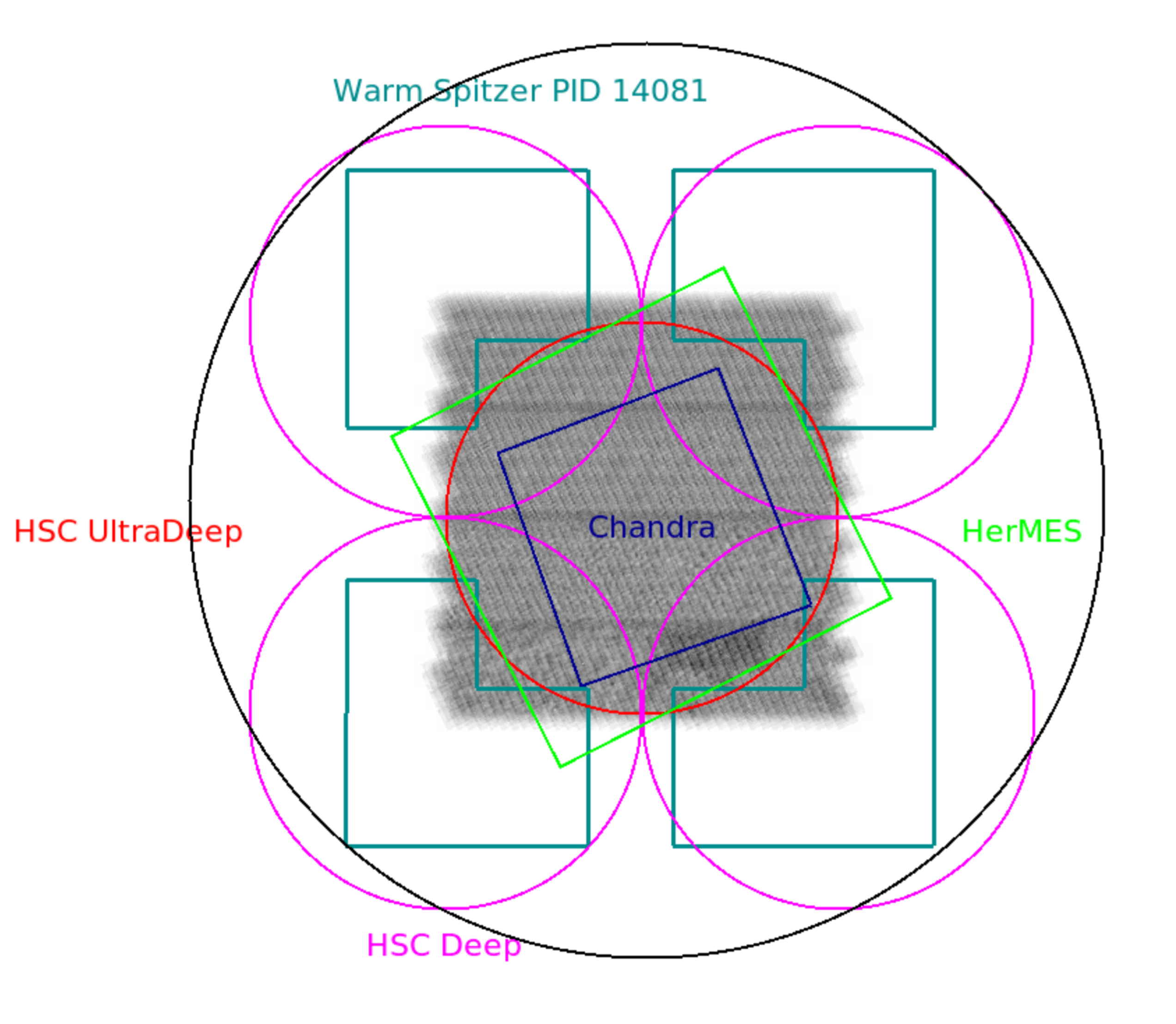,height=5.0truein,width=5.0truein,angle=0}
\vspace*{0.0in}
\caption[]{\protect\small 
Comparison of the planned LSST field vs.\ some of the key available 
multiwavelength data (as labeled) for COSMOS. The background grayscale
image shows the available \spitzer\ 3.6~$\mu$m data.}
\vspace*{-0.2in}
\end{figure}


\clearpage

\begin{figure}[t!]
\vskip -0.10truein
\hbox{
\hskip +0.4in 
\psfig{figure=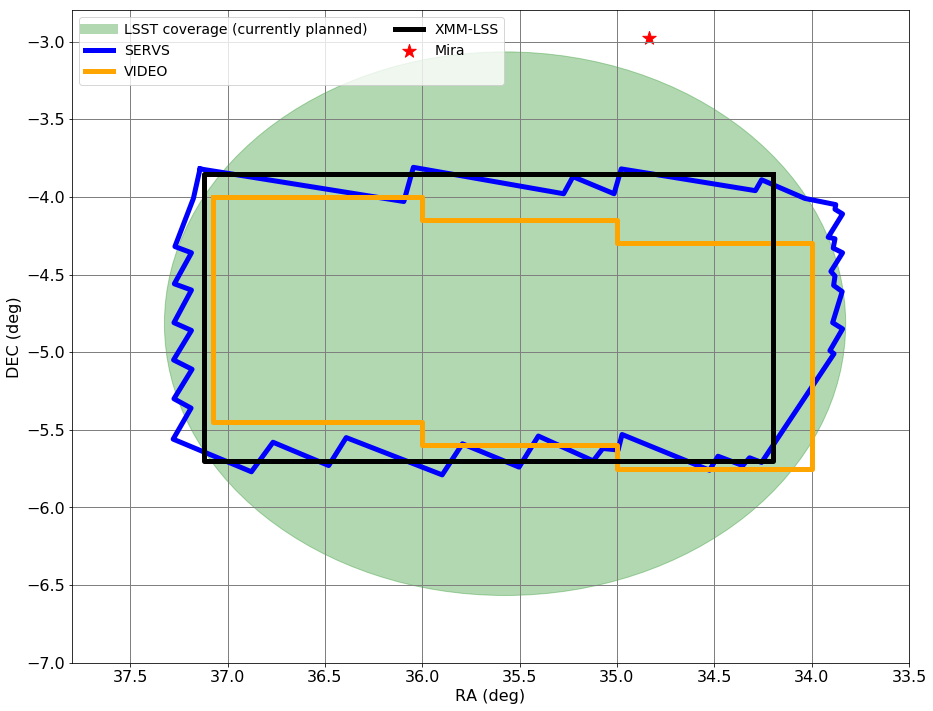,height=3.5truein,width=4.5truein,angle=0}
}
\vspace*{0.0in}
\caption[]{\protect\small 
The relative locations of the XMM-LSS field and the bright ($V=6.5$)
star Mira. An analysis of possible optical artifacts due to Mira is 
needed by LSST imaging-system experts. See \S3.2 for further discussion.}
\vspace*{-0.2in}
\end{figure}


\end{document}